\documentclass[preprint,amsmath,amssymb,11pt,aps,showpacs,showkeys]{revtex4}
\usepackage{graphicx}
\usepackage{graphics}
\usepackage{amsmath}
\usepackage{dcolumn}
\usepackage{amssymb}
\usepackage{bm}
\usepackage[latin1]{inputenc}

\begin{document}
\title{Noninertial effects on the ground state energy of a massive scalar field in the cosmic string spacetime}
\author{H. F. Mota}
\email{hm288@sussex.ac.uk}
\affiliation{Department of Physics and Astronomy, University of Sussex, Brighton, BN1 9QH, UK.}

\author{K. Bakke}
\email{kbakke@fisica.ufpb.br}
\affiliation{Departamento de F\'isica, Universidade Federal da Para\'iba, Caixa Postal 5008, 58051-970, Jo\~ao Pessoa, PB, Brazil.}

\begin{abstract}
We investigate the influence of noninertial effects on the ground state energy of a massive scalar field in the cosmic string spacetime. We generalize the results obtained by Khusnutdinov and Bordag [N. R. Khusnutdinov and M. Bordag, Phys. Rev. D {\bf59}, 064017 (1999)] to a noninertial reference frame and show, by contrast, that a non-vanishing contribution to the ground state energy stems from the noninertial effects. Moreover, we show that there is no influence of the curvature of the spacetime on this non-vanishing contribution to the ground state energy of a massive scalar field.  
\end{abstract}


\maketitle

\section{Introduction}

The Casimir effect has been a subject of many investigations since the first experimental test performed by Sparnaay \cite{casimir} in the late 1950's. This effect, which was predicted by H. Casimir \cite{HBG}, originally arises from quantum fluctuations in the vacuum of the electromagnetic field and, as a consequence, a finite vacuum energy (associated with an attractive force) appears induced by material boundaries in contrast to the vacuum energy defined in the Minkowski spacetime. Besides, from the experimental point of view, the reality of the effect has been extensively confirmed by many experiments \cite{GLK, AWR, GLK2}, which has opened a window for a variety of physical applications in different areas such as Biology and Chemistry \cite{VAP, VMM1}. 

Due to the quantum nature of the Casimir effect, it can also arise from the quantum fluctuation of other relativistic quantum fields, for instance, scalar and fermionic fields \cite{VBB1, VBB2}.  Furthermore, it has  also been realized that the Casimir effect can be investigated in the context of spacetimes with nontrivial topology, in which cases there is no need to have a material boundary by imposing boundary conditions, since the own topology of these backgrounds is responsible for imposing conditions on the considered fields \cite{VMM1, VMM}. A particular kind of spacetime with nontrivial topology is the cosmic string spacetime. The cosmic string is a linear topological defect which originally arises, in the context of some 
gauge field theories, as a consequence of a symmetry breaking phase transition in the early Universe. It can either form closed loops or extend to infinity \cite{AV}. A particular 
interest in the present paper is the latter case, which describes a spacetime with a conical topology. Although locally there is no gravity describing the cosmic string spacetime, there exist several interesting gravitational effects associated with the nontrivial topology of the spacelike region of the cosmic string spacetime. Among these effects, a cosmic string can act as a gravitational lens \cite{vil}, it can produce the Casimir effect \cite{JSD} and it can be studied as a background in other different contexts \cite{kibble,staro,kat,kleinert,moraesG2,furt,b,b2,b4,furt2,mello}.

Our interest in this brief report is to investigate the influence of noninertial effects on the ground state energy of a massive scalar field in the cosmic string
spacetime. The study of noninertial effects has discovered interesting effects in the context of quantum mechanics, as example, in interferometry associated with geometric phases \cite{sag,sag5,r3}, and a new coupling between the angular momentum and the angular velocity of the rotating frame \cite{r1,r2,r4}. Other discussions about phase shifts in the wave function of a quantum particle in noninertial systems have been made via Lorentz transformations \cite{r5}, in the weak field approximation \cite{r6}, and by obtaining the analogue effect of the Aharonov-Casher effect \cite{bf12}. Other studies of noninertial effects in quantum systems has also been extended to confined systems, such as, persistent currents in quantum rings \cite{r11}, spin currents \cite{r12}, Dirac fields \cite{r10}, scalar fields \cite{r8}, the Dirac oscillator \cite{b5}, rotational and gravitational effects in quantum interference \cite{r14,r15,r16} and the confinement of a neutral particle to a quantum dot \cite{b2,b4}. 

In recent years, Khusnutdinov and Bordag \cite{Nail} showed that the ground state energy of a massive scalar field in the cosmic string spacetime is zero. In this work, we generalize the results of Khusnutdinov and Bordag \cite{Nail} to a noninertial reference frame and show, by contrast, that noninertial effects yield a non-vanishing contribution to the ground state energy. Moreover, we show that this non-vanishing contribution to the ground state energy of a massive scalar field does not depend on the topology of the cosmic string spacetime.

This paper is organized as follows: in section II, we introduce the Klein-Gordon equation for a massive scalar field in a noninertial reference frame in the cosmic string spacetime, then, we discuss the influence of noninertial effects on the ground state energy of a massive scalar field; in section III, we present our conclusions.

\section{Massive Scalar Field in noninertial reference frame in the cosmic string spacetime}

In this section, we discuss the influence of the noninertial effects of a rotating reference frame on the eigenfrequencies of a massive scalar field in the cosmic string 
spacetime. We begin by writing the line element of the cosmic string spacetime and, in the following, we obtain the eigenfrequencies of the massive scalar field in a 
noninertial reference frame. The line element of the cosmic string spacetime is characterized by the presence of a parameter related to the deficit of angle, which is defined as 
$\alpha=1-4G\mu$, where $\mu$ is the dimensionless linear mass density of the cosmic string, $G$ is the gravitational Newton constant and the azimuthal angle is defined in 
the range: $0\leq\Phi<2\pi$. Working with the units $\hbar=c=1$, the line element of the cosmic string spacetime is given by \cite{vil,kibble}
\begin{eqnarray}
ds^{2}=-dT^{2}+dR^{2}+\alpha^{2}R^{2}d\Phi^{2}+dZ^{2}.
\label{2.1}
\end{eqnarray}

Furthermore, the geometry described by the line element (\ref{2.1}) possesses a conical singularity represented by the curvature tensor $R_{\rho,\varphi}^{\rho,\varphi}=\frac{1-\alpha}{4\alpha}\,\delta_{2}(\vec{r})$, where $\delta_{2}(\vec{r})$ is the two-dimensional delta function. This behaviour of the curvature tensor is denominated as a conical singularity \cite{staro}, which gives rise to the curvature concentrated on the cosmic string axis and possesses a null curvature in all other points. It is worth mentioning that all values of the parameter $\alpha\,>\,1$ correspond to a spacetime with negative curvature which does not make sense in the general relativity context \cite{kleinert,kat,furt}. Therefore, the parameter $\alpha$ given in the line element (\ref{2.1}) can only assume values for which $\alpha\,<\,1$. 

Now, let us make the following coordinate transformation: 
$T=t$, $R=\rho$, $\Phi=\varphi+\varpi\,t$ and $Z=z$, where $0\leq\varphi<2\pi$ and $\varpi$ is the constant angular velocity of the rotating frame. Thus, the line element (\ref{2.1}) becomes
\begin{eqnarray}
ds^{2}=-\left(1-\varpi^{2}\alpha^{2}\rho^{2}\right)\,dt^{2}+2\varpi\alpha^{2}\rho^{2}d\varphi\,dt+d\rho^{2}+\alpha^{2}\rho^{2}d\varphi^{2}+dz^{2}.
\label{2.3}
\end{eqnarray}

Hence, the line element (\ref{2.3}) describes a scenario of general relativity corresponding to the cosmic string spacetime background in a rotating coordinate system. We should note that the line element (\ref{2.3}) is defined in the range $0\,<\rho\,<\,1/\varpi\alpha$ \cite{landau3,b,b2,b4,b5}. Values where $\rho\,>\,1/\varpi\alpha$ mean that the line element (\ref{2.3}) is not well-defined because this region of the spacetime corresponds to a particle placed outside of the light-cone. This interesting restriction of the radial coordinate imposed by noninertial effects gives rise to a hard-wall confining potential, where the geometry of the spacetime plays this role of a hard-wall confining potential. In the context of quantum mechanics, if the wave function of the quantum particle cross the limit $\rho=1/\alpha\varpi$, thus, a non-null probability of finding the particle outside the light-cone exists and, as a consequence, the velocity of the particle would be greater than the velocity of the light. In this sense, a hard-wall confining potential stems from both noninertial effects and the topology of the cosmic string spacetime by imposing that the wave function of the quantum particle vanishes at $\rho\rightarrow1/\alpha\varpi$. Recently, the behaviour of the Dirac oscillator frequency has been analysed under the influence of noninertial effects and the topology of the cosmic string spacetime \cite{b5}. In the present work, we show that this restriction allow us to analyse the influence of noninertial effects and the topology of the cosmic string spacetime on the Casimir energy, that is, on the ground state energy which stems from the eigenfrequencies of a scalar field.

Now, let us consider a uncharged scalar quantum particle embedded in the scenario of general relativity described by the line element (\ref{2.3}). We deal with this system within the framework of general relativity, then, the field equation that describes this quantum dynamics is given by the Klein-Gordon equation in curved spacetime \cite{bd}. In this way, the Klein-Gordon equation becomes
\begin{eqnarray}
m^{2}\phi=-\frac{\partial^{2}\phi}{\partial t^{2}}+2\varpi\,\frac{\partial^{2}\phi}{\partial\varphi\,\partial t}-\varpi^{2}\,\frac{\partial^{2}\phi}{\partial\varphi^{2}}+\frac{\partial^{2}\phi}{\partial\rho^{2}}+\frac{1}{\rho}\,\frac{\partial\phi}{\partial\rho}+\frac{1}{\alpha^{2}\rho^{2}}\,\frac{\partial^{2}\phi}{\partial\varphi^{2}}+\frac{\partial^{2}\phi}{\partial z^{2}},
\label{2.4}
\end{eqnarray}
with $m$ being the mass of the scalar particle. We can observe that $\phi$ is an eigenfunction of the operators $\hat{p}_{z}=-i\partial_{z}$ and $\hat{L}_{z}=-i\partial_{\varphi}$, thus, we can write the solution of Eq. (\ref{2.4}) in terms of the eigenvalues of the operator $\hat{p}_{z}$ and $\hat{L}_{z}$ in the following form:
\begin{eqnarray}
\phi\left(t,\,\rho,\,\varphi,\,z\right)=e^{-i\,\omega\,t}\,e^{i\,l\,\varphi}\,e^{i\,k\,z}\,f\left(\rho\right),
\label{2.5}
\end{eqnarray}
where $\omega$ and $k$ are constants and $l=0,\pm1,\pm2,\ldots$. Then, Substituting the solution (\ref{2.5}) into Eq. (\ref{2.4}), we obtain the following radial equation,
\begin{eqnarray}
\frac{d^{2}f}{d\rho^{2}}+\frac{1}{\rho}\,\frac{df}{d\rho}-\frac{l^{2}}{\alpha^{2}\rho^{2}}\,f+\lambda^{2}\,f=0,
\label{2.6}
\end{eqnarray}
where we have defined the parameter $\lambda$ in the form:
\begin{eqnarray}
\lambda^{2}=\left(\omega+l\varpi\right)^{2}-m^{2}-k^{2}=\omega_{\mathrm{eff}}^{2}-m^{2}-k^{2},
\label{2.7}
\end{eqnarray}
where $\omega_{\mathrm{eff}}=\omega+l\,\varpi$. From now on, we consider just the positive values of the angular momentum $l$ in order to keep $\omega_{\mathrm{eff}}$ being positive definite, thus, we write  $\omega_{\mathrm{eff}}=\omega+\left|l\right|\,\varpi$. Note that, by writing $\omega_{\mathrm{eff}}=\omega+\left|l\right|\,\varpi$, we keep $\omega_{\mathrm{eff}}$ being positive definite and the values of the angular momentum quantum number defined as $l=0,\pm1,\pm2,\ldots$. Then, we have that Eq. (\ref{2.6}) is the Bessel differential equation \cite{abra}. The general solution of Eq. (\ref{2.7}) is given in the form: $f\left(\rho\right)=A\,J_{\frac{\left|l\right|}{\alpha}}\left(\lambda\rho\right)+B\,N_{\frac{\left|l\right|}{\alpha}}\left(\lambda\rho\right)$, where $J_{\frac{\left|l\right|}{\alpha}}\left(\lambda\rho\right)$ and $N_{\frac{\left|l\right|}{\alpha}}\left(\lambda\rho\right)$ are the Bessel function of first kind and second kind \cite{abra}. In order to have a regular solution at the origin, we must take $B=0$ in the general solution since the Neumann function diverges at the origin. Thus, the regular solution of Eq. (\ref{2.6}) at the origin is given by: 
\begin{eqnarray}
f\left(\rho\right)=A\,J_{\frac{\left|l\right|}{\alpha}}\left(\lambda\,\rho\right). 
\label{2.8}
\end{eqnarray}

Returning to our previous discussion about the restriction of the radial coordinate due to noninertial effects made in Eq. (\ref{2.3}), we have seen that all values of the radial coordinate given by $\rho\,>\,1/\alpha\,\varpi$ mean that the particle is situated outside the light-cone. In the context of the quantum field theory, this restriction of the values of the radial coordinate imposes that the scalar field must vanish at $\rho\rightarrow\rho_{0}=1/\alpha\varpi$, that is, the radial part of the scalar field (\ref{2.8}) must satisfy the boundary condition:
\begin{eqnarray}
f\left(\rho\rightarrow\rho_{0}=1/\alpha\varpi\right)=0.
\label{2.9}
\end{eqnarray}
This means that the geometry of the spacetime plays the role of a hard-wall confining potential \cite{b,b2,b4,b5}. Thereby, from Eq. (\ref{2.8}), we have $J_{\frac{\left|l\right|}{\alpha}}\left(\lambda\,\rho_{0}\right)=0$. The boundary condition (\ref{2.9}) implies that $\lambda\rightarrow\lambda_{n,\,\,l}$, where $n=1,2,3,\ldots$ correspond to the set of zeros of the Bessel function. Therefore, by using Eq. (\ref{2.7}), we can establish the relation between the eigenfrequencies and the angular velocity of the rotating frame:
\begin{eqnarray}
\omega_{n,\,\,l}=\left(\lambda_{n,\,\,l}^{2}+m^{2}+k^{2}\right)^{1/2}-\left|l\right|\varpi.
\label{2.10}
\end{eqnarray}
The last term of Eq. (\ref{2.10}) corresponds to the coupling between the angular velocity and the angular momentum quantum number, which is a Sagnac-type effect \cite{sag,r4}. It is worth mentioning that, in the nonrelativistic limit, this coupling between the angular velocity and the angular momentum quantum number is known as the Page-Werner {\it et al.} term \cite{r1,r2}. On the other hand, the first term of Eq. (\ref{2.10}) corresponds to the eigenfrequencies associated with the topology of the cosmic string spacetime.

Henceforth, let us discuss the ground state energy of this system from the eigenfrequencies obtained in Eq. (\ref{2.10}). The ground state energy of the field is given by 
(with the units $\hbar=c=1$)
\begin{eqnarray}
E_{0}=\frac{1}{2}\sum_{n=1}^{\infty}\,\sum_{l=-\infty}^{\infty}\omega_{n,\,\,l}=\frac{1}{2}\sum_{n=1}^{\infty}\left[\sum_{l=-\infty}^{\infty}\left(\lambda_{n,\,\,l}^{2}+m^{2}+k^{2}\right)^{1/2}-\sum_{l=-\infty}^{\infty}\left|l\right|\varpi\right].
\label{2.11}
\end{eqnarray}

For this purpose, one can analyse each part of Eq. (\ref{2.11}) separately. The first term of Eq. (\ref{2.11}) corresponds to the renormalized ground state energy of a massive scalar field in the cosmic string spacetime, which was analysed by Khusnutdinov and Bordag \cite{Nail} by using the zeta functional regularization procedure. The interesting result obtained by Khusnutdinov and Bordag \cite{Nail} is that the renormalized ground state energy in this case is zero. 

On the other hand, the last term of Eq. (\ref{2.11}) which stems from noninertial effects yields a non-vanishing contribution to the ground state energy given by
\begin{eqnarray}
\bar{E}_{0}=-\frac{1}{2}\sum_{n=1}^{\infty}\,\sum_{l=-\infty}^{\infty}\left|l\right|\varpi=-\varpi\,\sum_{n=1}^{\infty}\,\sum_{l=1}^{\infty}\,l.
\label{2.12}
\end{eqnarray}

Observe that the sum given in Eq. (\ref{2.12}) diverges, therefore, we need to apply a regularization procedure in order to obtain a finite energy. For this purpose, we can make use of the properties of the Riemann zeta function \cite{abra}
\begin{eqnarray}
\zeta(s)=\sum_{k=1}^{\infty}k^{-s}.
\label{2.13}
\end{eqnarray}
Thereby, the Riemann zeta function (\ref{2.13}) is calculated in Eq. (\ref{2.12}) for the cases $s=0$ and $s=-1$, which corresponds to the sum over the indices $n$ and $l$, respectively. In this way, the sums in Eq. (\ref{2.12}) yield a renormalized energy given by $E_{0}^{\mathrm{ren}}=-\varpi\,\zeta(0)\zeta(-1)$ and, due to $\zeta(0)=-1/2$ and $\zeta(-1)=-1/12$, thus, the renormalized energy associated with noninertial effects is
\begin{eqnarray}
E_{0}^{\mathrm{ren}}=-\frac{\varpi}{24}.
\label{2.14}
\end{eqnarray}

Hence, in contrast to the results of Khusnutdinov and Bordag \cite{Nail}, we have that noninertial effects yield a finite contribution to the ground state energy of a scalar field in the cosmic string background. Moreover, this finite contribution to the ground state energy does not depend on the topology of the cosmic string spacetime. Observe that we can recover the result of Ref. \cite{Nail} by taking the limit $\varpi\rightarrow 0$.

In recent decades, the influence of curvature of a disclination (corresponding to the spatial part of the line element of the cosmic string spacetime) on the vacuum polarization of the electromagnetic field was investigated in Ref. \cite{moraes}. Based on the results of Ref. \cite{moraes}, an interesting point of discussion can be the contribution of noninertial effects and curvature to the vacuum polarization of the electromagnetic field. Another point of interest can be the influence of torsion and noninertial effects on the Casimir energy. The contribution of torsion to the Casimir energy was studied in Ref. \cite{moraes2} by considering a topological defect called screw dislocation. Moreover, it has been shown in Ref. \cite{moraes3} that torsion effects can modify the electromagnetic field, therefore, new contributions of torsion and noninertial effects to the vacuum polarization of the electromagnetic field can be expected.

\section{conclusions}

In this brief report, we have investigated the ground state energy of a massive scalar field in the cosmic string spacetime by considering the presence of noninertial effects. We have seen that noninertial effects restrict the radial coordinate to a range given by $0\,<\rho\,<\,1/\varpi\alpha$ and, thus, where the massive scalar field can be defined. In this way, we can interpret this restriction of the physical region of the spacetime imposed by noninertial effects as giving rise to a hard-wall confining potential, whose geometry of the spacetime can play the role of a hard-wall confining potential. 

Thereby, we have solved the Klein-Gordon equation in the general relativity background defined by the line element (\ref{2.3}) by imposing the Dirichlet boundary condition (\ref{2.9}). As a consequence, the noninertial effects yield an additional term to the eigenfrequencies of the scalar field given by $-|l|\varpi$, which corresponds to the coupling between the angular momentum and the angular velocity \cite{r4}. Therefore, we can interpret the eigenfrequencies (\ref{2.10}) as consisting in a term associated with the topology of the cosmic string spacetime and a term which arises from the coupling between the angular momentum and the angular velocity. 

Moreover, by analysing the ground state energy of the scalar field, we have seen that noninertial effects yield a finite contribution to the ground state energy of the scalar field in the cosmic string background in contrast to the results of Ref. \cite{Nail}, whose contribution to the ground state energy of a scalar field from the topology of the cosmic string spacetime is zero. Besides, we have shown that the finite contribution to the ground state energy obtained in Eq. (\ref{2.14}) does not depend on the topology of the cosmic string spacetime, that is, it depends only on the noninertial effects. By taking the limit $\varpi\rightarrow 0$, we have also shown that the results of Ref. \cite{Nail} are recovered.

\acknowledgments

The authors would like to thank E. R. Bezerra de Mello for interesting discussions and the Brazilian agencies CNPq and CAPES for financial support.

\end{document}